\documentclass[twocolumn,amsmath,showpacs,aps,prb,floatfix]{revtex4-1}

\usepackage{graphicx}
\usepackage{amssymb}
\usepackage{hyperref}

\begin{document}

\begin{onecolumngrid}
\scriptsize
\noindent
{\tt The following article appeared in J.\,Vac.\,Sci.\,Technol.\,B {\bf 30}, 03D112 (2012) and may be found at \href{http://dx.doi.org/10.1116/1.4706892}{http://link.aip.org/link/?jvb/30/03D112}.}
\end{onecolumngrid}
\vspace{0.34 in}

\title{Valley and spin polarization from graphene line defect scattering}

\author{Daniel~Gunlycke}
\affiliation{Naval Research Laboratory, Washington, D.C.~20375, USA}
\author{Carter~T.~White}
\affiliation{Naval Research Laboratory, Washington, D.C.~20375, USA}

\begin{abstract}
Quantum transport calculations describing electron scattering off an extended line defect in graphene are presented.  The calculations include potentials from local magnetic moments recently predicted to exist on sites adjacent to the line defect.  The transmission probability is derived and expressed as a function of valley, spin, and angle of incidence of an electron at the Fermi level being scattered.  It is shown that the previously predicted valley polarization in a beam of transmitted electrons is not significantly influenced by the presence of the magnetic moments.  These moments, however, do introduce some spin polarization, in addition to the valley polarization, albeit no more than about 20\%.
\end{abstract}


\maketitle

\section{Introduction}

\begin{figure}[h!]
    \includegraphics{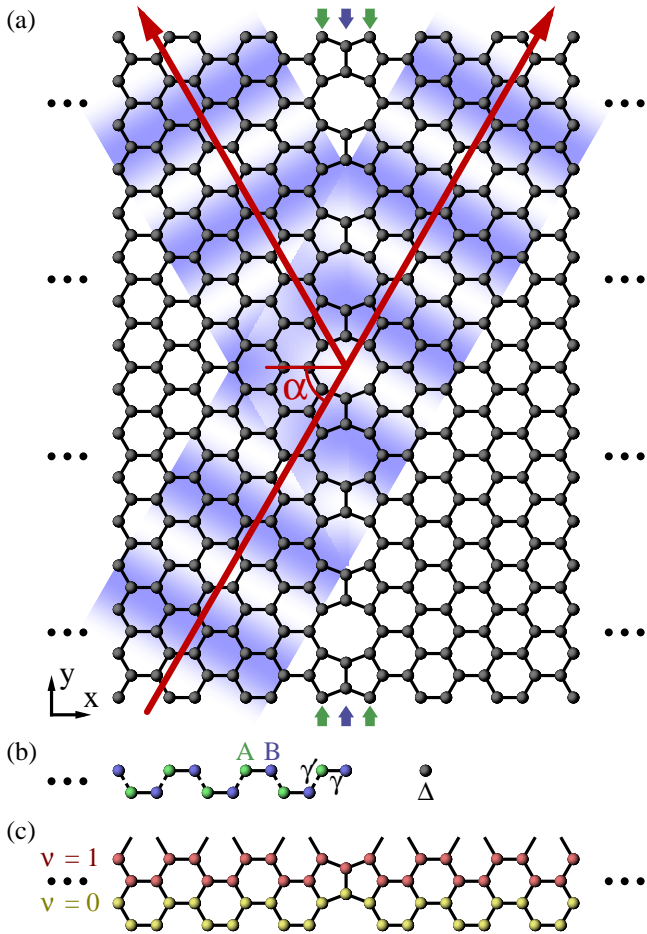}
    \caption{Extended line defect in graphene.  (a) An electronic Bloch wave approaching the line defect at an angle of incidence $\alpha$ is being scattered.  This scattering is influenced by local magnetic moments on the sites (between green arrows) next to the line defect sites (between blue arrows).  (b) The semi-infinite sheet of graphene to the left of the line defect transformed into momentum space along the line defect.  The chain has alternating couplings, $\gamma$ and $\gamma'$, making the self energy $\Delta$ representing the influence of the sites away from the end point depend on whether the end site is of type A (green) or B (blue).  (c) Primitive cell of the graphene line defect.  The sites are paired to form the two chains $\nu=0,1$.}
    \label{f.1}
\end{figure}
The success of electronics rests on the ability to control electron motion.  Such control is typically achieved by varying the electrostatic potential in a semiconductor with a suitable band gap.  This straightforward way to control electron motion has proven tremendously successful and is the main reason for the considerable effort by the graphene community devoted to graphene nanoribbons\cite{Klei94_1,Fuji96_1,Han07_1} and bilayer graphene in the presence of an electric field.\cite{Lu06_1,Guin06_1,McCa06_2,Zhan09_1}  Another more subtle way to control electron motion is though scattering off deliberate defects in the material.\cite{Yazy10a,Gunl11_1}  This approach could add new functionality and ultimately prove to be the future for electronics.  Graphene is a promising material for controlled electron scattering for several reasons:  (i) it has a well-defined structure,\cite{Wall47_1,Novo05_1} (ii) owing to its sp$^2$ hybridization, it has $\pi$-orbitals near the Fermi level that can form extended states,\cite{Wall47_1} (iii) it is a semi-metal with only a limited number of scattering channels available near the Fermi level,\cite{Wall47_1} and (iv) it offers an electron mobility\cite{Bolo08_1,Du08_1,Orli08_1} high enough to support ballistic transport in the micron range.\cite{Bolo08_1}

Herein, we consider electron scattering off the extended line defect in graphene illustrated in Fig.\,\ref{f.1}(a).  This structure both preserves the sp$^2$ hybridization of carbon and is precisely defined.  It is not a hypothetical structure, but one that has already been observed in experiments.\cite{Lahi10}  An arbitrary electron scattering off the line defect occupies a state that away from the line defect approaches asymptotically one of graphene, and if the energy of this electron is near the graphene Fermi level, it can in addition to its energy be identified by its direction of motion, its valley, and its spin.  Herein, a tight-binding model is used to derive the transmission probability of an arbitrary incident electron.  The model includes a potential to describe ferromagnetically aligned local magnetic moments that has been shown to be present in the line defect structure.\cite{Whit12_1}  These moments break the spin-degeneracy, otherwise present, causing there to be spin polarization among the electrons of a transmitted beam.  This spin polarization is found to be rather small, limiting its use.  More important is that the local magnetic moments do not appear to degrade the predicted valley polarization of the transmitted electrons near the Fermi level.\cite{Gunl11_1}  Therefore, the graphene line defect remain an illustrative example of a system where the valley degree of freedom can be exploited instead of or alongside the spin degree of freedom for applications in quantum information processing.

The next section develops the theoretical formalism to describe the electron scattering off the line defect.  This theory is applied in Sec.\,\ref{s.3} to the scattering of electrons near and at the Fermi level.  Conclusions drawn from the results are presented in Sec.\,\ref{s.4}, including symmetry argument explaining the large valley polarization for electrons scattering at high angles of incidence.

\section{Scattering formalism}

Our objective is to obtain the transmission probability for an arbitrary right-moving electron approaching the line defect.  The scattering calculations are founded on a tight-binding model with a basis set consisting of orthonormal $\pi$-orbitals.\cite{Wall47_1}  Herein, we consider nearest-neighbor interactions with a hopping parameter $\gamma=-2.6$\,eV.  Longer-range interactions\cite{Gunl11_1} and distortions\cite{Jian11} do not qualitatively change the results and has therefore been ignored for presentational clarity.

The scattering problem is solved in steps.  First, we recognize that the structure in Fig.\,\ref{f.1}(a) can be viewed as a set of line defect sites connected to two semi-infinite graphene sheets.  Next, we solve for the self energy representing all interactions within one of the semi-infinite sheets.  This can be achieved by exploiting translational symmetry along the $y$-direction.  Once the self energy has been derived, we can then calculate the retarded Green function on the line defect sites, which is needed to obtain the sought after transmission probability.

To keep the notation tidy, much of the formalism below is presented in dimensionless units, which can be recognized by their diacritic tildes.  Energy units are scaled with the absolute value of the hopping parameter $\gamma$.  Other dimensionless units are also introduced below, as needed.

\subsection{Semi-infinite graphene}

The influence of sites away from the line defect is captured by a self energy defined at the edges of the semi-infinite sheets of graphene on each side of the line defect.  To derive this self energy, we first recognize that a semi-infinite sheet of graphene with a zigzag edge has translational symmetry in the direction along the edge with a period equal to the graphene lattice constant $a$.  If a Bloch wave is considered with an arbitrary wave vector $k_y$ along the direction of translational symmetry, we can transform the semi-infinite graphene sheet into a semi-infinite linear chain with alternating couplings $\gamma$ and $\gamma'\equiv2\gamma\cos\tilde{k}_y$, where we have defined $\tilde{k}_y\equiv k_ya/2$.  See Fig.\,\ref{f.1}(b).  Self-energy recurrence relations can be generated through a process of replacing the sites away from the end site with a self energy, adding a site to the chain, and recalculate the self energy at the new end site.  In this case, the recurrence relations become
\begin{align}
	\tilde{\Delta}^A&=4\cos^2\tilde{k}_y\left(\tilde{E}-\tilde{\Delta}^B\right)^{-1}\\
	\tilde{\Delta}^B&=\left(\tilde{E}-\tilde{\Delta}^A\right)^{-1},
	\label{e.1}
\end{align}
where $\tilde{E}$ is the energy of the Bloch wave measured relative to the energy at Fermi level and $\tilde{\Delta}^\lambda$ is the self energy for an end site of type $\lambda\in\{A,B\}$.  The end site type is related to the sublattice of the semi-infinite graphene sheet to which the edge sites belong.  The retarded solution to the self-energy recurrence relations can be expressed as
\begin{align}
	\tilde{\Delta}^A&=\frac{1}{\tilde{E}}\Big[4\cos^2\tilde{k}_y+2\cos\tilde{k}_y\,e^{i\tilde{k}_x}\Big]\\
	\tilde{\Delta}^B&=\frac{1}{\tilde{E}}\Big[1+2\cos\tilde{k}_y\,e^{i\tilde{k}_x}\Big],
	\label{e.2}
\end{align}
where
\begin{align}
	\tilde{k}_x&=\pi+\operatorname{sgn}\tilde{E}\left[\pi-\operatorname{Re}\left\{\arccos\frac{\tilde{E}^2-1-4\cos^2\tilde{k}_y}{4\cos\tilde{k}_y}\right\}\right]\nonumber\\
	&+i\left|\operatorname{Im}\left\{\arccos\frac{\tilde{E}^2-1-4\cos^2\tilde{k}_y}{4\cos\tilde{k}_y}\right\}\right|.
	\label{e.3}
\end{align}

The local moments on the sites adjacent to the line defect sites (between the green arrows) can be modeled through a spin-dependent onsite potential $\tilde{\varepsilon}_\sigma$, where $\sigma\in\{-1,1\}$ is the spin.  Assuming the Hubbard parameter $\tilde{U}=1.06$ [\onlinecite{Gunl07_4}] and a difference of the average spin population $\langle n_{\sigma}\rangle$ per semi-infinite chain of $\langle n_1\rangle-\langle n_{-1}\rangle=1/6$, we approximate the onsite potential to be $\tilde{\varepsilon}_\sigma=\tilde{U}\big[\langle n_{-\sigma}\rangle-\langle n_{\sigma}\rangle\big]/2\approx\mp0.09$ in dimensionless energy units for spin $\sigma=\pm1$, respectively.\cite{Whit12_1}  The local moments affect the self energy with the end point at the line defect, which is given by
\begin{equation}
	\tilde{\Delta}=\left(\tilde{E}-\tilde{\varepsilon}_\sigma-\tilde{\Delta}^A\right)^{-1}.
	\label{e.4}
\end{equation}

There is translation symmetry, not only in the semi-infinite graphene sheets, but also in the full line defect structure in Fig.\,\ref{f.1}(a).  The line defect structure has a period $2a$ along the line defect, i.e., twice the period of the semi-infinite graphene sheet.  The primitive cell of the line defect structure is shown in Fig.\,\ref{f.1}(c), where the sites have been divided into bottom, $\nu=0$, and top, $\nu=1$, sites.  Rather than using the real space basis $|\nu\rangle$, it is more convenient to use a basis $|n\rangle$, in which the self energy is diagonal.  To find this basis, we first recognize that the wave vector of the line defect structure must be conserved in the scattering process.  As a result, the Bloch wave in the semi-infinite graphene sheet with wave vector $k_y$ can only couple to one other Bloch wave in the scattering process, the one with wave vector $k_y+\pi/a$.  From the phase relationship between equivalent sites with $\nu=0,1$ imposed by the translational symmetry of the semi-infinite graphene sheet, we obtain
\begin{equation}
	|n\rangle = \frac{1}{\sqrt{2}}\sum_\nu e^{i(2\tilde{k}_y+n\pi)\nu}|\nu\rangle,
	\label{e.5}
\end{equation}
where $n=0,1$.  As the states $|n\rangle$ are eigenstates of the self-energy operator $\tilde{\Sigma}$, we have
\begin{equation}
	\langle n|\tilde{\Sigma}|n'\rangle=\tilde{\Sigma}_n\delta_{nn'},
	\label{e.6}
\end{equation}
where $\tilde{\Sigma}_n$ is the self energy $\tilde{\Delta}$ in Eq.\,(\ref{e.4}) calculated for the Bloch wave with wave vector $k_y+n\pi/a$.

In calculating the transmission probability, we also need the elements of the broadening operator $\tilde{\Gamma}\equiv i\left(\tilde{\Sigma}-\tilde{\Sigma}^\dagger\right)$.  From this definition, we see that $|n\rangle$ are also eigenstates of $\tilde{\Gamma}$, yielding the elements
\begin{equation}
	\langle n|\tilde{\Gamma}|n'\rangle=-2\operatorname{Im}\tilde{\Sigma}_n\,\delta_{nn'}.
	\label{e.7}
\end{equation}

\subsection{Line Defect}

Let us focus on the center two sites in the primitive cell in Fig.\,\ref{f.1}(c) forming the line defect.  With the coupling of these sites to all other sites in the primitive cell already accounted for through the self energy, the only coupling in the Hamiltonian is the coupling $\gamma$ between the two sites with $\nu=0$ and $\nu=1$.  In the basis set defined by Eq.\,(\ref{e.5}), the Hamiltonian elements are
\begin{equation}
	\langle n|\tilde{H}|n'\rangle=-\frac{1}{2}\left[e^{i(2\tilde{k}_y+n'\pi)}+e^{-i(2\tilde{k}_y+n\pi)}\right].
	\label{e.8}
\end{equation}
Given the self energy and the Hamiltonian, we can calculate the retarded Green function operator $\tilde{G}=\big(\tilde{E}I-\tilde{H}-2\tilde{\Sigma}\big)^{-1}$, where $I$ is the unit operator.  The factor 2 in front of the self energy operator reflects the connections of the line defect to the two semi-infinite graphene sheets.  Using Eq.\,(\ref{e.6}) and Eq.\,(\ref{e.8}), we find the elements of the retarded Green function operator, which can be expressed as
\begin{align}
	\langle n|\tilde{G}|n'\rangle&=\left\{
	\begin{array}{cc}
		\frac{\tilde{E}-\cos(2\tilde{k}_y+n\pi)-2\tilde{\Sigma}_{1-n}}{\det\big(\tilde{E}I-\tilde{H}-2\tilde{\Sigma}\big)} & \quad\mathrm{for}~n=n',\vspace{0.05 in}\\
		\frac{i\sin(2\tilde{k}_y+n\pi)}{\det\big(\tilde{E}I-\tilde{H}-2\tilde{\Sigma}\big)} & \quad\mathrm{for}~n\ne n',
	\end{array}
	\right.
	\label{e.9}
\end{align}
where
\begin{align}
	\det\big(\tilde{E}I-\tilde{H}-2\tilde{\Sigma}\big)=\,&\big(\tilde{E}-2\tilde{\Sigma}_0\big)\big(\tilde{E}-2\tilde{\Sigma}_1\big)-1\nonumber\\
	&+2\cos2\tilde{k}_y\big(\tilde{\Sigma}_0-\tilde{\Sigma}_1\big).
	\label{e.10}
\end{align}
Using the elements of the broadening operator in Eq.\,(\ref{e.7}) and the Green function operator in Eq.\,(\ref{e.10}), we find that the probability that a state $|n\rangle$ transmits through the line defect into state $|n'\rangle$ is given by
\begin{align}
	T_{n\rightarrow n'}&=\langle n|\tilde{\Gamma}|n\rangle\langle n|\tilde{G}|n'\rangle\langle n'|\tilde{\Gamma}|n'\rangle\langle n'|\tilde{G}^\dagger|n\rangle\nonumber\\
	&=4\operatorname{Im}\tilde{\Sigma}_n\operatorname{Im}\tilde{\Sigma}_{n'}\Big|\langle n|\tilde{G}|n'\rangle\Big|^2.
	\label{e.11}
\end{align}

\section{Electrons at the Fermi level}
\label{s.3}

\begin{figure}
    \includegraphics{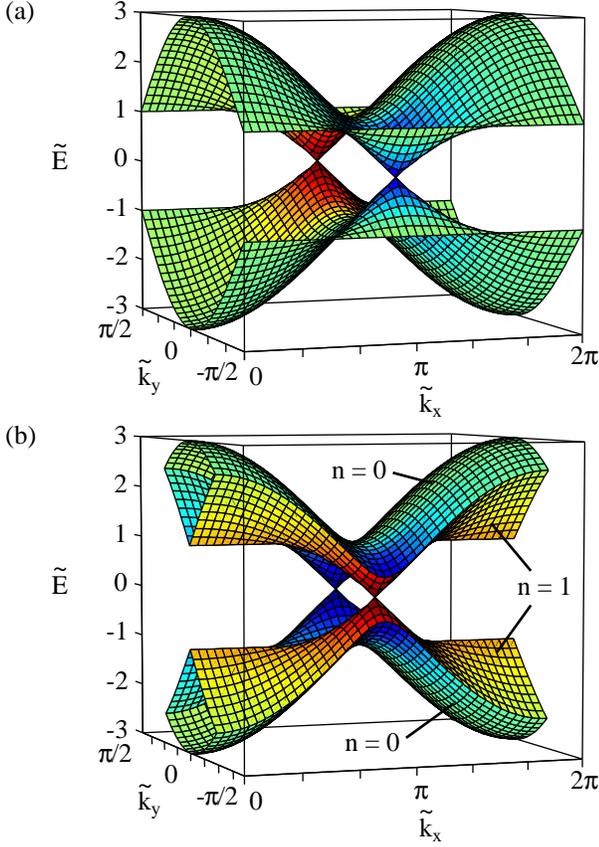}
    \caption{Energy of the asymptotic states given by the graphene band structure.  (a) The band structure presented using a rectangular reciprocal primitive cell, in which the the two valley $\tau=\pm1$ are shown as blue and red, respectively.  The energy is set to zero at the Fermi level.  (b) Graphene band structure folded to be commensurate with the line defect structure.  Bands farthest and closest to the Fermi level have $n=0$ and $n=1$, respectively.}
    \label{f.2}
\end{figure}

Because the scattering process conserves energy and the wave vector associated with the translation symmetry along the line defect, it is natural to develop the scattering formalism based on these parameters.  Although these parameters together with the index $n$ form a parameter space covering all possible asymptotic graphene states, there are other more intuitive representations.  An arbitrary graphene state is typically described by a spin, a band index, and a two-dimensional wave vector $\vec{k}=\left(k_x,k_y\right)$.  Rather than using the conventional hexagonal first Brillouin zone in graphene, it is herein more convenient to use an alternative reciprocal primitive cell bounded by $k_x\in\left[0,4\pi/\sqrt{3}a\right[$ and $k_y\in\left[-\pi/a,\pi/a\right[$.  Assuming an extended right-moving asymptotic graphene state, the wave vector component $k_x$ is determined from Eq.\,(\ref{e.3}) with $\tilde{k}_x\equiv\sqrt{3}k_xa/2$.  The graphene band structure, given by\cite{Wall47_1}
\begin{equation}
	\tilde{E}=\eta\sqrt{1+4\cos^2\tilde{k}_y+4\cos\tilde{k}_y\cos\tilde{k}_x},
	\label{e.12}
\end{equation}
where $\eta=\pm1$ refers to the conduction and valence bands, is shown in Fig.\,\ref{f.2}(a).
To make the asymptotic graphene state commensurate with the line defect structure, we fold the graphene band structure as shown in Fig.\,\ref{f.2}(b), where in this case $k_y\in\left[-\pi/2a,\pi/2a\right[$.  The bands furthest and closest to the Fermi level, located where the Dirac cones meet, correspond to $n=0$ and $n=1$, respectively.

\begin{figure}
    \includegraphics{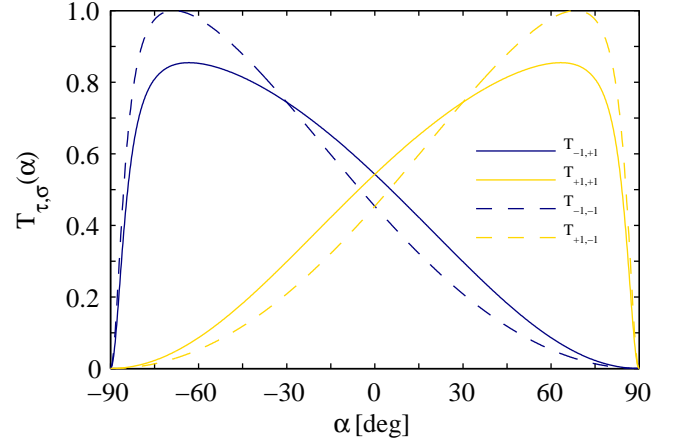}
    \caption{Transmission probability of an incident electron at the Fermi level in valley $\tau$ and with spin $\sigma$, approaching the line defect at the angle of incidence $\alpha$.}
    \label{f.3}
\end{figure}

Let us focus our attention to the asymptotic states of most interest, i.e. those near the Fermi level.  For energies with $|\tilde{E}|<\sqrt{2}-1$, which include all energies within an eV from the Fermi level, $\operatorname{Im}\tilde{\Sigma}_0=0$, i.e. all extended states are in the $n=1$ band.  In this energy regime, elastic scattering does not permit any interband scattering at the line defect, and thus the transmission probability $T_{\tau,\sigma}$, given by $T_{1\rightarrow1}$ in Eq.\,(\ref{e.11}), is
\begin{equation}
	T_{\tau,\sigma}=\frac{4\Big(\operatorname{Im}\tilde{\Sigma}_1\Big)^2\left(\tilde{E}+\cos2\tilde{k}_y-2\tilde{\Sigma}_0\right)^2}{\left|\big(\tilde{E}-2\tilde{\Sigma}_0\big)\big(\tilde{E}-2\tilde{\Sigma}_1\big)-1+2\cos2\tilde{k}_y\big(\tilde{\Sigma}_0-\tilde{\Sigma}_1\big)\right|^2}.
	\label{e.13}
\end{equation}
The expression above is an exact result that can be evaluated numerically.  To make further analytical progress, we focus on the low energy limit.  In doing so, we first introduce the graphene wave vector $\vec{q}=(q_x,q_y)$, where $q_x = k_x-2\pi/\sqrt{3}a$ and $q_y = k_y+2\pi\tau/3a$, centered at valley $\tau\in\{-1,1\}$.  In our low-energy regime, we can use $\tau$, $\sigma$, and $\vec{q}$ to describe the asymptotic graphene state.  To lowest order in $q\equiv|\vec{q}|$, the energy dispersion in Eq.\,(\ref{e.12}) is $E=\eta\hbar v_Fq$, where $v_F=\sqrt{3}|\gamma|a/2\hbar$ is the Fermi velocity.  Next, we introduce the angle of incidence $\alpha$ shown in Fig.\,\ref{f.1}(a).  From the group velocity relation $\tan\alpha=\big(\partial E/\partial q_y\big)\Big/\big(\partial E/\partial q_x\big)=q_y\big/ q_x$ and the assumption of a right-moving state, which gives $\operatorname{sgn} q_x=\eta$, we obtain $q_x=(E/\hbar v_F)\cos\alpha$ and $q_y=(E/\hbar v_F)\sin\alpha$.  Therefore, the asymptotic graphene state can be expressed uniquely by the energy $E$, valley $\tau$, spin $\sigma$, and angle of incidence $\alpha$ of the incident electron.  After expressing the transmission probability in Eq.\,(\ref{e.13}) using these quantities, we find the zero energy limit
\begin{equation}
	T_{\tau,\sigma}(\alpha)=\left|\frac{\operatorname{Im}\tilde{\Sigma}_1}{1+\tilde{\Sigma}_1}\right|^2=\frac{1}{1+\left[\frac{1-\sin\tau\alpha+\tilde{\varepsilon}_\sigma\big(1-2\sin\tau\alpha\big)+\tilde{\varepsilon}_\sigma^2}{\cos\tau\alpha}\right]^2}
	\label{e.14}
\end{equation}
From this expression, shown in Fig.\,\ref{f.3}, we see that $T_{\tau,\sigma}(\alpha)=T_\sigma(\tau\alpha)$, which implies that changing valley has the same effect as changing the sign of the angle of incidence.
By letting $\partial T_\sigma\big/\partial\tau\alpha = 0$, we find that the transmission has a stationary point at
\begin{equation}
	\tau\alpha=\arcsin\left(\frac{1+2\tilde{\varepsilon}_\sigma}{1+\tilde{\varepsilon}_\sigma+\tilde{\varepsilon}_\sigma^2}\right)^\sigma.
	\label{e.15}
\end{equation}
Inserted into Eq.\,(\ref{e.14}), this relation gives the maximum transmission
\begin{equation}
	T_\sigma^\mathrm{max}=\left\{
	\begin{array}{cc}
		\left(1-2\tilde{\varepsilon}_\sigma-\tilde{\varepsilon}_\sigma^2+2\tilde{\varepsilon}_\sigma^3+\tilde{\varepsilon}_\sigma^4\right)^{-1} & \quad\sigma=+1,\vspace{0.05 in}\\
		1 & \quad\sigma=-1,
	\end{array}
	\right.
	\label{e.16}
\end{equation}
This maximum at the stationary point can be seen in Fig.\,\ref{f.3}.

If, rather than a single electron, a beam of electrons is sent towards the line defect, the scattered electrons would be both valley- and spin-polarized.
\begin{figure}[h!]
    \includegraphics{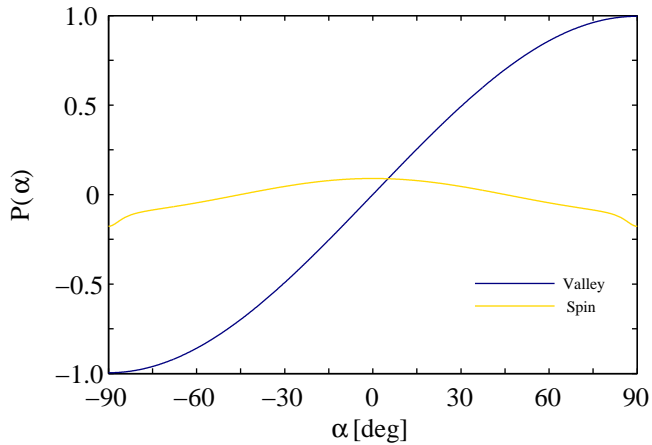}
    \caption{Spin and valley polarization of the transmitted portion of a beam of electrons at the Fermi level after scattering off the line defect at the angle of incidence $\alpha$.}
    \label{f.4}
\end{figure}
The former valley polarization, defined as
\begin{equation}
	\mathcal{P}_\mathrm{v}\equiv\frac{\sum_\sigma T_{1,\sigma}-\sum_\sigma T_{-1,\sigma}}{\sum_{\tau\sigma}T_{\tau,\sigma}},
	\label{e.17}
\end{equation}
is shown in Fig.\,\ref{f.4}.  This polarization is very close to $\mathcal{P}_\mathrm{v}=\sin\alpha$ predicted previously in the absence of the local magnetic moments.\cite{Gunl11_1}  Similarly, we can define the spin polarization of the beam of transmitted electrons, as
\begin{equation}
	\mathcal{P}_\mathrm{s}\equiv\frac{\sum_\tau T_{\tau,1}-\sum_\tau T_{\tau,-1}}{\sum_{\tau\sigma}T_{\tau,\sigma}}.
	\label{e.18}
\end{equation}
Although there is some spin polarization, as can be seen in Fig.\,\ref{f.4}, this polarization is small compared to the valley polarization.

\section{Conclusions}
\label{s.4}
Graphene is a promising material for controlling electron motion through scattering off well-defined defects.  Herein, we have shown that electrons near the Fermi level scattered off an observed extended line defect can be both valley- and spin-polarized.  The spin polarization, arising from local magnetic moments on sites adjacent to the line defect, is found to be less than 20\%.  The valley polarization, on the other hand, can reach near 100\%.

The valley filtering taking place at the line defect, which is very different from other suggested methods for obtaining valley polarization,\cite{Ryce07,Tkac09,Zhai10,Wu11} can be understood from its reflection symmetry.  Consider an asymptotic graphene state near the Fermi level, which could be expressed as $|\Phi_\tau\rangle=\left(|A\rangle+ie^{-i\theta}|B\rangle\right)/\sqrt{2}$, where $|A\rangle$ and $|B\rangle$ refer to the two graphene sublattices and $\theta$ is a pseudospin angle providing the phase relationship between the two sublattices.  The only asymptotic graphene states also eigenstates of the reflection operator, which maps the $A$ sublattice on one side of the line defect onto the $B$ sublattice on the opposite side, and vice versa, are those with $\theta=\pm\pi/2$.  These states are symmetric and antisymmetric, respectively.  Antisymmetric states must have a node at the line defect, making the coupling across the line defect, and concomitantly the transmission, small.  Although, the potential describing the local magnetic moments is a source of scattering, the transmission through the symmetric states is generally good.  As a result, the line defect allows Bloch waves with $\theta=\pi/2$ to transmit, while blocking those with $\theta=-\pi/2$.

The asymptotic graphene state $|\Phi_\tau\rangle$ is, in general, not an eigenstate of the reflection operator and has $\theta\ne\pm\pi/2$.  It can, however, always be written as a superposition of the symmetric and antisymmetric states.  If the transmission through the symmetric and antisymmetric states are 1 and 0, respectively, the transmission probability can be obtained from the modulus square of the symmetric component of the incident graphene state, i.e. $(1+\sin\theta)/2$.  From the graphene eigenstates, it can also be shown that the pseudospin angle $\theta=\tau\alpha$, yielding a transmission probability equal to that in Eq.\,(\ref{e.14}) in the absence of the potential describing the local magnetic moments.\cite{Gunl11_1}

To summarize, the valley filtering is a consequence of the imbalance between the transmission probabilities for the symmetric  and antisymmetric components of the incident graphene state.  This imbalance originates from the symmetry of the line defect structure.  As neither the introduction of longer-range interactions,\cite{Gunl11_1} distortion,\cite{Jian11} or potentials from the presence of local magnetic moments, considered herein, can undo the imbalance, we conclude that the valley filtering is a robust property of the graphene line defect for high angles of incidence.


\begin{acknowledgments}
The authors acknowledge support from the U.S. Office of Naval Research, directly and through the U.S. Naval Research Laboratory.
\end{acknowledgments}


%


\end{document}